\documentclass{PoS}
\newcommand{\be}{\begin{equation}}
\newcommand{\ee}{\end{equation}}

\title{Cosmic-ray heating of molecular cloud cores}

\ShortTitle{Cosmic-ray heating of molecular cloud cores}

\author{\speaker{Daniele Galli} \\
INAF-Osservatorio Astrofisico di Arcetri, Largo E. Fermi 5, 50125 Firenze, Italy \\
E-mail: \email{galli@arcetri.astro.it}
}
\author{Marco Padovani~$^{a,b}$\\
\llap{$^{a}$} Laboratoire Univers et Particules de Montpellier, UMR 5299 du CNRS, Universit\'e de Montpellier II, place E. Bataillon, cc072, 34095 Montpellier, France\\
\llap{$^{b}$} INAF--Osservatorio Astrofisico di Arcetri, Largo E. Fermi 5, 50125, Firenze, Italy\\
E-mail: \email{Marco.Padovani@lupm.univ-montp2.fr}}

\abstract{Cosmic rays are an important source of heating in the
interstellar medium, in particular in dense molecular cloud cores shielded
from the external ultraviolet radiation field.
The limits placed on the cosmic-ray ionization rate from measurements of 
the gas temperature in dense  clouds are unaffected by
the uncertainties associated
to the traditional methods based on the analysis of molecular
abundances. However, high-resolution data are required to determine
with sufficient accuracy the spatial temperature distribution within prestellar cores. In
this contribution we illustrate in detail the case of the well-studied
prestellar core L1544, showing that both its thermal structure and
chemical composition are consistent with a cosmic ray ionization
rate of $\sim 10^{-17}$~s$^{-1}$, significantly smaller than the
value measured in the diffuse interstellar medium. We also briefly discuss
possible applications of this method to the molecular clouds of
other galaxies.}

\FullConference{Cosmic Rays and the InterStellar Medium\\
		 24-27 June 2014\\
		 Montpellier, France}

\begin{document}

\section{Introduction}

\begin{figure}
\vspace{-4cm}
\begin{center}
\includegraphics[width=.7\textwidth]{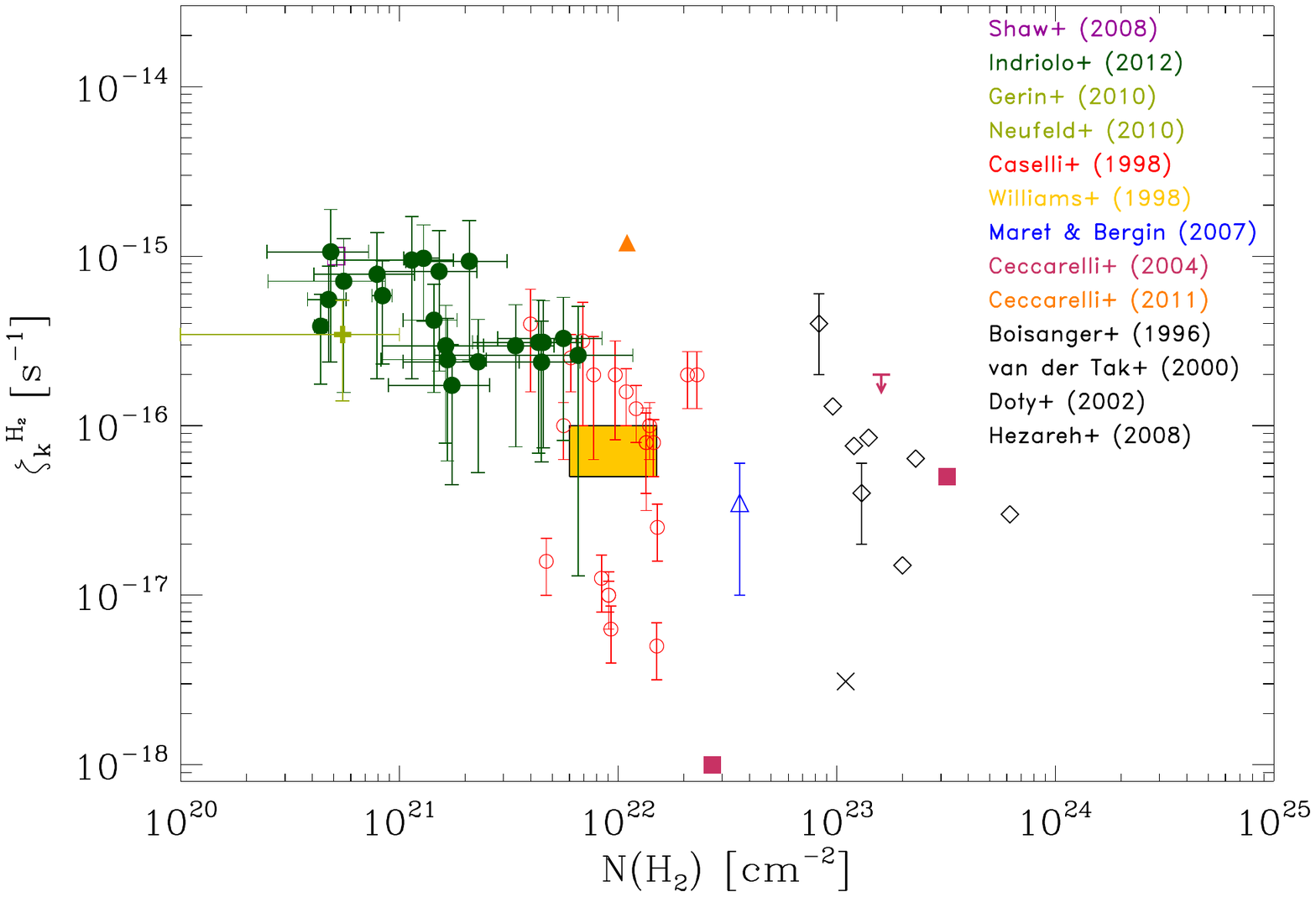}
\end{center}
\vspace{-4cm}
\caption{Values of the CR ionization rate of H$_2$ in diffuse clouds (green dots), molecular cloud cores (red circles)
and massive protostellar envelopes (black diamonds) as function of the cloud's column
density. For references, see Padovani et al.~(2009).}
\label{zetacr}
\end{figure}

The recent discovery of significant amounts of H$_3^+$ in diffuse
molecular clouds (e.g., Indriolo et al.~2007, 2009 and references therein)
has spurred a significant increase of interest in the study of the
propagation and ionization rate of low-energy cosmic rays (hereafter
CRs) in the interstellar medium.  A straightforward analysis of the
observations of Indriolo et al. (2007) extended by Indriolo \&
McCall (2012) to a sample of 50 diffuse molecular clouds, yields
values of the ionization rate $\zeta$ for H$_2$ in the range $\sim
(2-11) \times 10^{-16}$~s$^{-1}$, significantly larger than the
``standard'' value $\zeta\approx 10^{-17}$~s$^{-1}$ that had been
in use since the pioneering works of Hayakawa et al.~(1961) and Spitzer
\& Tomasko~(1968). This unexpected result has led to the reconsideration
of how CR protons and electrons propagate in the interstellar
medium (e.g., Indriolo et al.~2009, Padovani et al.~2009, Padovani
\& Galli~2011, Everett \& Zweibel~2011, Rimmer et al.~2012).
CR ionization rates have also been measured in dense molecular clouds and
massive stellar envelopes, by modeling the observed abundance of molecular 
ions like HCO$^+$ and DCO$^+$. The values of $\zeta$ in these environments
show considerable scatter, either intrinsic or due to
uncertainties in the chemical modeling of the primary data. In any
case the average value of $\zeta$ in dense molecular clouds is
between one or two orders of magnitude smaller than in diffuse
clouds. Fig.~\ref{zetacr} shows a summary of the observational determination
of $\zeta$ in clouds of column density $N({\rm H}_2)$ ranging from $10^{20}$~cm$^{-2}$ to 
$10^{24}$~cm$^{-2}$. In addition to a considerable scatter, the data suggest a general trend
of $\zeta$ decreasing with increasing $N({\rm H}_2)$, as expected in simple 1-D models
of CR attenuation (Padovani et al.~2009). Understanding the dependence of $\zeta$ on the physical
characteristics of the environment (density, column density, magnetic fields, etc.)
is a fundamental step for modelling the ionization fraction in the interstellar medium
and the coupling of the gas with the magnetic field. This, in turn, has important consequences on 
theoretical models of cloud collapse and disk formation (Padovani et al. 2103, 2014).

\section{Cosmic-ray heating}

In addition to ionization, another process relating to the interaction
of CRs with the interstellar medium is their role in heating the
atomic and the molecular gas.  CRs are an efficient (often dominant)
source of heating in various environments, from the dense gas in
molecular clouds (Goldsmith \& Langer~1978), both in normal and
starbust galaxies (Suchkov et al.~1993), to photodissociation regions
(Shaw et al.~2009), and possibly even in the primordial gas (Jasche
et al.~2007).

Considering only molecular clouds in our Galaxy, the available observational data are consistent with a picture in
which the dust temperature of prestellar cores (i.e. cores without 
any internal stellar energy source) in general decreases
toward the center (Ward-Thompson et al.~2002, Bianchi et al.~2003).
Observations in the mid-IR (Bacmann et al.~2000) and in the
far-IR (Ward-Thompson et al.~2002) suggest dust temperature gradients
consistent with heating from the external interstellar radiation
field. Recent results obtained by the sub-mm satellite
Herschel (e.g. Palmeirim et al.~2013) have confirmed these trends with a high degree 
of accuracy. On the theoretical side, available radiative transfer models
predict a factor of $\sim 2$ increase in dust temperature from
center to edge (Zucconi et al.~2001, Evans et al.~2001, Stematellos \& Whitworth~2003), 
with the gradient dependent on the cloud structure. 

Additional observational constraints on the thermal structure of
prestellar cores are given by spatially resolved measurements of
the gas temperature that has been accurately traced by interferometric
observations of molecular emission (in particular of NH$_3$) in
prestellar cores (Crapsi et al.~2007, Pagani et al.~2007) and dark
globules (Pineda \& Bensch~2007).  Low-mass cores show fairly uniform
gas temperature (e.g. Tafalla et al.~2002 for L1517B and L1498),
whereas massive quiescent cores in Orion show significant temperature
drops from edge to center (Li et al.~2003), as predicted by theoretical
models of dense, UV-shielded interstellar clouds (Falgarone \& Puget~1985,
Galli et al.~2002).  Thus, measurements of the
gas temperature in prestellar cores can be used to constrain the CR ionization
rate if radiative transfer models are also able to predict
the gas temperature distribution resulting from the balance of the
relevant heating and cooling mechanisms.

In general, the dust and gas temperatures $T_{\rm d}$ and $T_{\rm g}$ 
in a molecular cloud can be
computed by solving simultaneously the equations of thermal balance
of the gas and the dust
\be
\Gamma_{\rm ext}=\Lambda_{\rm d}(T_{\rm d})-\Lambda_{\rm gd}(T_{\rm d}, T_{\rm g}),
\ee
and 
\be
\Gamma_{\rm CR}=\Lambda_{\rm g}(T_{\rm g})+\Lambda_{\rm gd}(T_{\rm d}, T_{\rm g}),
\label{balance}
\ee
where $\Gamma_{\rm ext}$ is the dust heating rate per unit volume
from the external radiation field, $\Lambda_{\rm d}$ is the dust cooling rate
by infrared emission, $\Gamma_{\rm CR}$ is the CR
heating rate of the gas,  $\Lambda_{\rm g}$ the gas cooling rate by
molecular and atomic transitions, and $\Lambda_{\rm gd}$ the gas-dust
energy transfer rate.  Notice that we have ignored processes such
as the photoelectric heating since we are interested in regions
shielded by the external UV radiation field.  For a detailed
discussion of the heating and cooling functions that enter in the
thermal balance equations see e.g. Goldsmith (2001) and Galli et
al. (2002).  

\begin{figure}
\vspace{-2.5cm}
\begin{center}
\includegraphics[width=.7\textwidth]{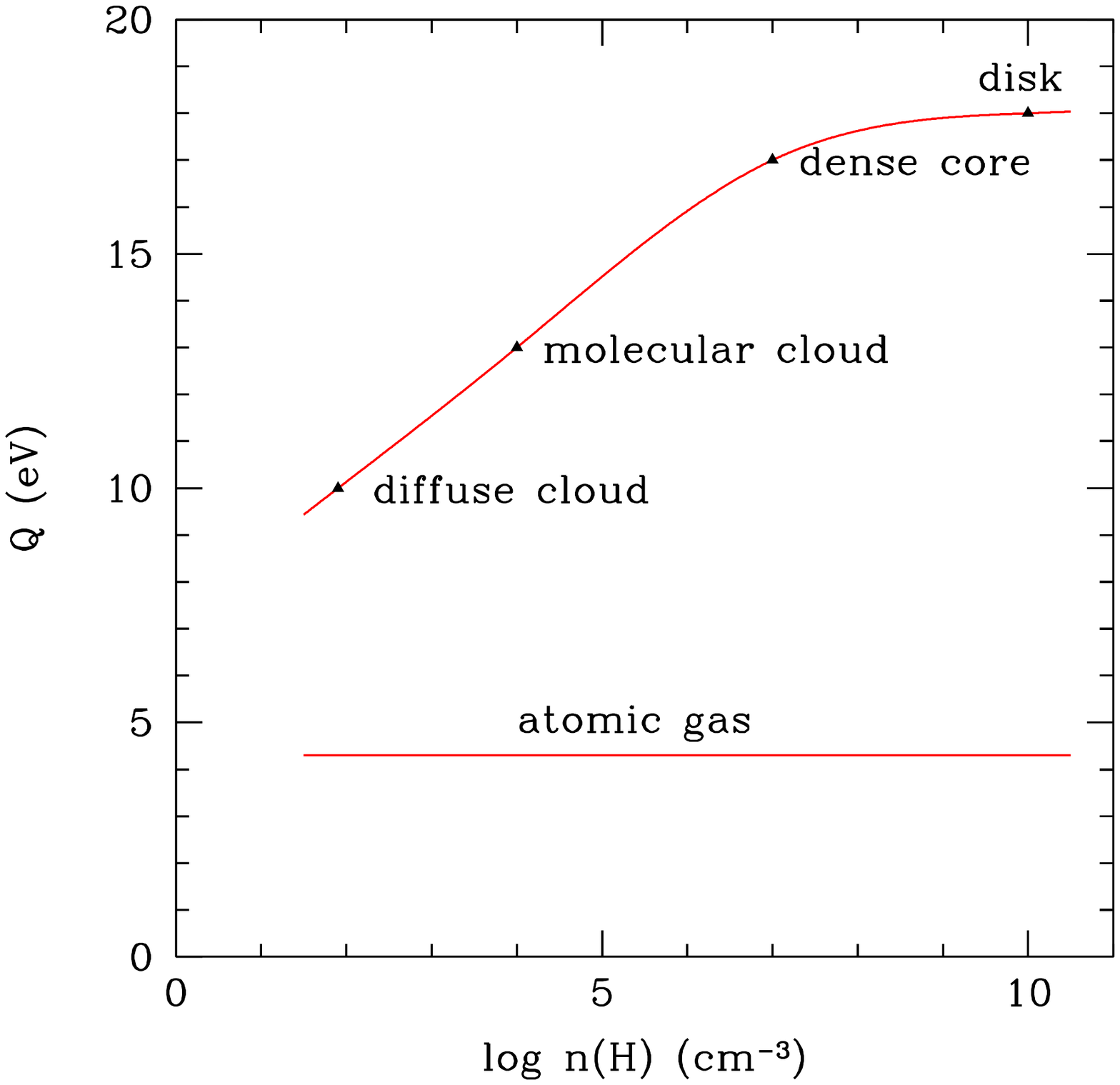}
\end{center}
\vspace{-4cm}
\caption{Average heat input per ionization $Q$ as function of the 
density of the environment for various astrophysical conditions (from 
Glassgold et al.~2012). Here $n({\rm H})$ the density of hydrogen in all
forms.}

\label{qcr}
\end{figure}

The CR heating $\Gamma_{\rm CR}$ is usually parametrized as
\be
\Gamma_{\rm CR}=n({\rm H}_2)\zeta Q,
\ee
where $n({\rm H}_2)$ is the H$_2$ density and $Q$ is the {\it mean} heat input per ionization.
Estimates of $Q$ available in the literature range over a factor
of three (Glassgold \& Langer~1973a,b, Cravens et al.~1975, 
Cravens \& Dalgarno~1978, Goldsmith \& Langer~1978, Goldsmith 2001).
Early studies suffered from
the poorly known electron cross sections in the early 70s, the
crude estimates of the energy-loss functions, and ignored the 
roles of H$^+$ and He$^+$
ions in molecular gas were ignored. A more up to date and complete
analysis was carried out by Dalgarno et al.~(1999) who considered
carefully all of the energy loss channels for electron energies up
to 1~keV in various mixtures of H, H$_2$ and He. They showed how
the energy expended to make an ion pair is partitioned among
elastic and several non-elastic processes, but they did not fully
treat the heating. This was accomplished by Glassgold et al.~(2012), who
used the results of Dalgarno et al.~(1999) to compute $Q$ with an
accuracy of $\sim 20$\% in a mixture of H$_2$ (or H) and He for
various astrophysical conditions (diffuse clouds, molecular clouds,
dense molecular cloud cores and protostellar disks). As shown by
Glassgold et al.~(2012), in dense molecular regions about 50\% of the energy of the ejected
electron can go into heating.  In addition,
CRs also produce ions and excited molecules that can interact with
the dominant neutral atomic or molecular gas. The products of these
reactions deposit in the gas a significant amount of the available
energy in the form of chemical heating, that represents a significant 
part of the CR heating. The average heat input per ionization (including the chemical heating), 
computed by Glassgold et al.~(2012) is shown schematically in Fig.~\ref{qcr}

Notice that a similar process arises in the X-ray irradiation of
molecular regions. In fact, X-ray and CR ionization are closely
related because the energy of the photon is almost entirely converted
into energy of the primary photoelectron, and therefore the
interactions of photons or CR electrons and nuclei with dense gas
are largely determined by the many fast supra-thermal electrons
they produce. Of course there are differences: X-ray ionization is
accompanied by fast Auger electrons as well as secondaries, and
CRs have an electron as well a nuclear component.

\section{A specific example: L1544}

As a specific example, we compare the results of the modelling
described in the previous section with the gas temperature profile
inferred from NH$_3$ observations of the starless core L1544, a
well-studied low-mass core in Taurus, which presents a large central
density ($n_{\rm c}({\rm H}_2) >10^6$~cm$^{-3}$).  Several observational
characteristics make this core a good candidate for being on the
point of becoming unstable (Crapsi et al. 2005) and collapse to form stars.  

To model the
thermal structure of L1544, we adopt the parameters derived by Galli
et al.~(2002) to match the thermal dust emission map obtained by
Ward-Thompson et al.~(1999).
To compute the gas temperature, we notice that the energy deposited
in the gas by CR ionization is negligible when compared to the
energy absorbed by the dust, so that the energy transfer between
gas and dust will not significantly affect the grain temperature.
Therefore, it is only necessary to solve the equation of thermal
equilibrium of the gas. According to these models, in the inner part of L1544 (densities
above $10^5$~cm$^{-3}$ ), the gas temperature is coupled to that
of the dust and thus decreases gradually toward the core center,
where $T_{\rm g}\approx T_{\rm d} \approx 6$~K.  For $\zeta
>10^{-18}$~s$^{-1}$ the gas is slightly hotter than the dust due
to CR heating and this difference increases as the density decreases.
In outer regions where the density is below $\sim 10^5$~cm$^{-3}$
and where shielding from the external radiation field is much less,
the gas and dust temperatures become uncoupled and one may find gas
temperatures higher or lower than that of the dust, depending on
the value of the CR ionization rate.  

\begin{figure}
\vspace{-2.5cm}
\begin{center}
\includegraphics[width=.7\textwidth]{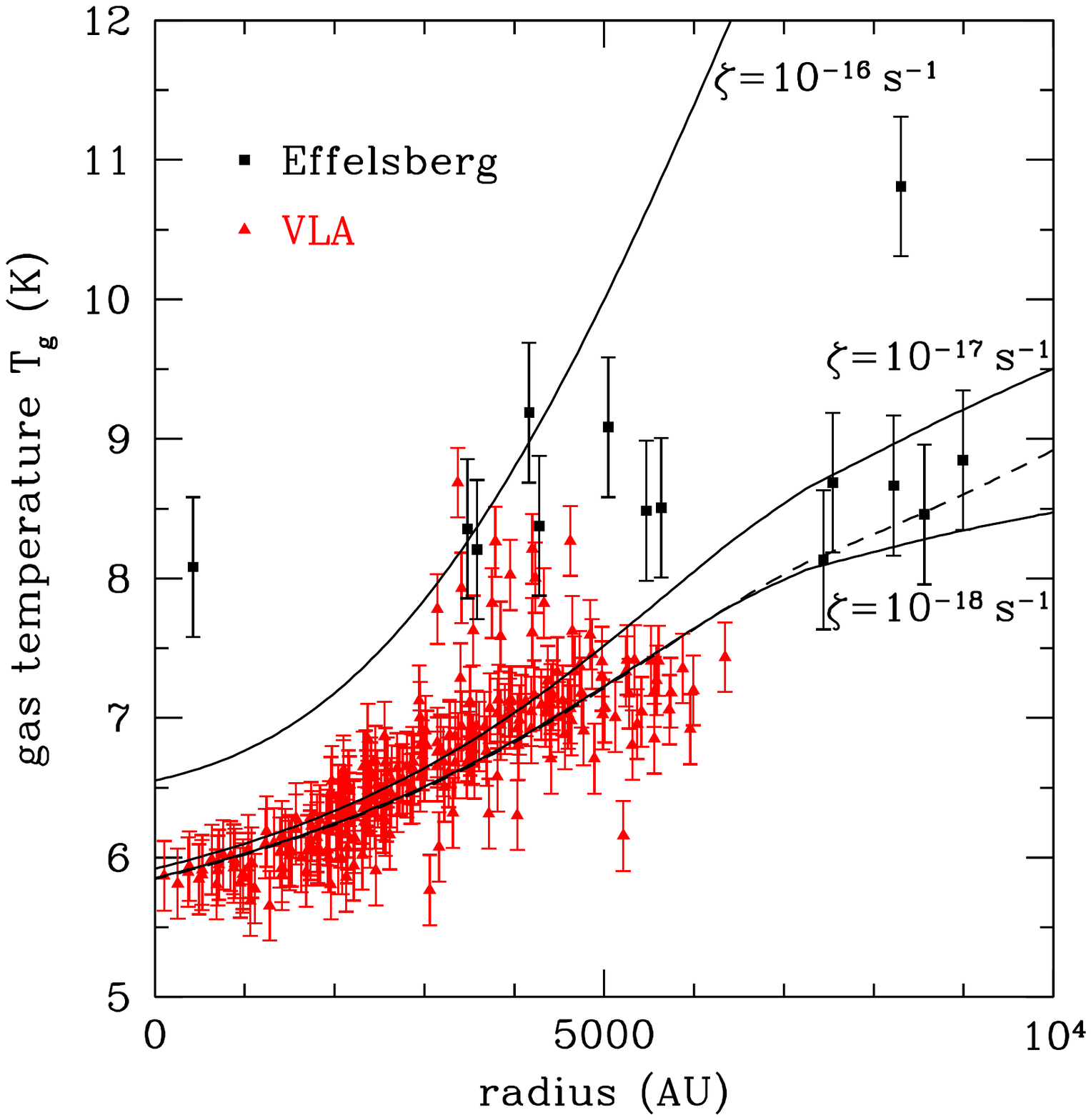}
\end{center}
\vspace{-4cm}
\caption{Thermal structure of the prestellar core L1544. The curves
show the radial profile of the gas and dust temperature computed
with the model of Galli et al.~(2002) for different values of the CR
ionization rate. From bottom to top: $\zeta=10^{-18}$~s$^{-1}$,
$10^{-17}$~s$^{-1}$, and $10^{-16}$~s$^{-1}$ (solid curves, gas
temperature; dashed curve, dust temperature).  The data show the
gas temperature derived from NH$_3$ observations by Tafalla et
al.~(2002) and Crapsi et al.~(2007) (black and red points,
respectively).}
\label{L1544}
\end{figure}

Fig.~\ref{L1544} shows the
dust and gas temperature of our model, whereas the black and red
points are the gas temperature determined from NH$_3$ by Tafalla
et al.~(2002) with the Effelsberg single-dish radiotelescope and
by Crapsi et al.~(2007) with the VLA interferometer array, respectively.
Since the models are spherically symmetric, the model temperature
profiles are shown as a function of radius, whereas the data are
plotted as a function of the projected distance from the dust peak.
The interferometer data constrain the gas temperature in the inner
regions of the core, showing a clear evidence of a temperature
gradient from $\sim 6$~K at the core's center to $\sim 8$~K at a
distance of $\sim 5000$~AU from the dust peak. The observed gradient
is in agreement with theoretical predictions if the CR ionization
rate is $\zeta \approx 10^{-17}$~s$^{-1}$ (or lower).  Notice that
a value of $\zeta = 10^{-16}$~s$^{-1}$, typical of diffuse molecular
clouds, results in a significantly higher gas temperature profile,
and is inconsistent with the VLA data.  These results compare
favourably with the analysis of the chemical abundances of L1544:
a best-fit chemical model reproducing the observed molecular column
densities at the dust peak and the observed abundance profiles gives
$\zeta=1.3 \times 10^{-17}$~s$^{-1}$ (Vastel et al.~2006).

\section{Discussion and conclusions}

The example illustrated in the previous section shows how measurements
of gas temperature in molecular cloud cores can be used to infer
or constrain the CR ionization rate.  This method may
represent an alternative, or a complement, to the traditional
approach based on the determination of abundance ratios of molecular
ions (Guelin et al.~1977, Caselli 1998), that suffers from limitations
associated to uncertainties in the chemical networks for dense
clouds.  Of course, as shown by the case of L1544, high-quality and
high-resolution data are needed to determine with sufficient accuracy
the gas temperature profile in molecular cloud cores (compare the
Effelsberg and VLA data in Fig.~\ref{L1544}). In addition, the
method is applicable to clouds (or cloud regions) of intermediate
density: if the density is too low (say for clouds of visual extinction
$A_V$ less than about 4), the heating of the gas is dominated by
the photoelectric effect on dust grains (see e.g. Le Bourlot et al.~1993),
whereas for densities larger than $\sim 10^6$ the gas is thermally
coupled to the dust. In either case, in these regimes the gas temperature becomes
insensitive to the local CR ionization rate.

The possibility of using measurements of the gas temperature to
infer the CR ionization rate has already been tested on molecular
clouds of external galaxies. In fact, heating by CRs has been invoked
to explain the presence of hot H$_2$ in the Galactic center and in
starburst galaxies (G\"usten et al.~1985, Lo et al.~1987).  In
particular, the temperature of molecular clouds in the starburst
galaxy M82, $T_{\rm g}\approx 50$--150~K, is compatible with a CR
ionization rate $\zeta\approx  4\times 10^{-15}$~s$^{-1}$ (Suchkov
et al.~1993), much in excess of the Galactic values discussed in
Section~1.  This CR enhancement, resulting from the boosted supernova
rate typical of a starburst galaxy, in turn implies a synchrotron
emission in M82 about $500$ times larger than in the disk of the
Milky Way, in agreement with radio observations (Seaquist et al.~1985)
and recent measurements of high-energy gamma-rays in M82 (VERITAS
Collaboration et al. 2009). The idea here is that the nuclear
component of CRs is responsible for heating the molecular gas and
producing the gamma-ray emission, while the associated low-energy
electronic component (primary and/or secondary) is responsible for
the synchrotron radio emission.

\acknowledgments
DG and MP acknowledge the support of the CNRS-INAF PICS project ``Pulsar wind nebulae,
supernova remnants and the origin of cosmic rays''.
MP also acknowledges the support of the OCEVU Labex (ANR-11-LABX-0060) 
and the A*MIDEX project (ANR-11-IDEX-0001-02) funded by the ``Investissements d'Avenir'' 
French government programme managed by the ANR.

\end{document}